# Field Ordering and Energy Density in Texture Cosmology

Nicholas G. Phillips [1]
*Department of Physics, University of Maryland, College Park, Maryland 20742*

A. Kogut [2]
*Hughes STX Corporation, Laboratory for Astronomy and Solar Physics, Code 685, NASA/GSFC, Greenbelt, MD 20771*

## ABSTRACT

We use numerical simulations of the time evolution of global textures to investigate the relationship between ordering dynamics and energy density in an expanding universe. Events in which individual textures become fully wound are rare. The energy density is dominated by the more numerous partially wound configurations, with median topological charge $\bar{\alpha}_\rho \simeq 0.44$. This verifies the recent supposition [11] that such partially wound configurations should dominate the cosmic microwave background.

[1] Electronic Address: phillips@zwicky.gsfc.nasa.gov
[2] Electronic Address: kogut@stars.gsfc.nasa.gov

# I. Introduction

Recent observations [1, 2] of anisotropy in the cosmic microwave background (CMB) provide strong evidence for primordial density perturbations. Several models predict the statistical distribution of these perturbations. In inflationary models, density perturbations result from parametric amplification of quantum fluctuations during the inflationary epoch [3], while in topological defect models the density perturbations result from large-scale ordering effects following a phase transition with broken symmetry [4, 5, 6, 7].

Observational tests of these models are complicated by the fact that cosmological models predict the CMB anisotropy to be a single realization of a stochastic process whose properties (power spectrum, phase correlations, etc.) are predicted only for an ensemble average of equivalent realizations. Any single realization may be expected to vary about the ensemble average. Although a large body of work exists providing analytic derivation of ensemble average properties of inflationary models, the non-linear nature of defect models preclude exact analytic results. Tests of defect models must use either time-consuming numerical simulations or an approximate analytic model.

Global textures are one class of defect models for which an approximate analytic model has been proposed [7]; however, the predictions of this approximate model do not agree with the results of a complete simulation [8]. One of the main assumptions of this model is that the CMB anisotropy is dominated by the most energetic isolated defects. We use a numerical simulation of the complete theory to investigate this assumption and find that the dominant contribution comes instead from the far more numerous but less energetic defects. This was first suggested in [11] for a non-expanding universe. By doing the simulation for an expanding universe, we demonstrate that the less energetic defects must be considered in any model of the texture generated CMB anisotropy.

Topological defects such as monopoles, domain walls, or cosmic strings involve regions of space where the order field remains in the higher-energy symmetric state. The energy density in these models is dominated by the order field potential. Textures [6] are a class of topological defects with the property that the broken symmetric phase vacuum manifold of the order field is of the same dimension as the spatial geometry. This allows the order field always to stay on the vacuum manifold, regardless of initial conditions. Perturbations in the energy density are driven solely by the order field's kinetic energy.

The topological charge of a region of space is defined as the fraction of the vacuum manifold covered as the order field is mapped into the manifold [9]. At the time of the phase transition, the order field in each causal domain will randomly fall into any of the degenerate vacua. As the different domains come into causal contact, the field may develop a non-trivial topological charge, forming a wound configuration or "knot". Derrick's theorem [10] states that the spatial extent of any region of non-zero gradient energy must shrink, causing the order field to become increasingly tightly wound around the vacuum manifold. The kinetic energy in this wound configuration will eventually become greater than the potential barrier to the symmetric state, lifting the order field from the vacuum manifold and allowing the texture to "unwind" as it reduces the topological charge.

Since this unwinding process is so energetic, it is often viewed as the dominant contribution to the energy density. Turok & Spergel [7] provide an analytic model of the CMB anisotropy resulting from a distribution of unwinding spherical knots. The results of this model are not reproduced by numerical simulations [8]. On the other hand, Borrill *et. al.* have pointed out that unwinding events are rare: they found only 40 such events in 1000 simulations [9]. This agrees with analytic predictions [6] (see also [4, 15]). They also find partially-wound texture configurations can still generate anisotropies similar to those generated by events which do unwind [11].

We have simulated the order field dynamics in the non-linear $\sigma$ model approximation to investigate the distribution of gradient energy density. We identify both the wound and partially-wound events in our simulation. The unwinding events are indeed rare: we find only 3 unwindings events out of over 400 texture events. Although unwindings are the most energetic individual events, most of the energy density is associated with the more numerous partially-wound texture configurations.

## II. Analysis

Our calculations are done in a spatial flat Robertson-Walker homogeneous universe, with metric given by

$$ds^2 = -dt^2 + a(t)^2(dx^2 + dy^2 + dz^2) = a(\tau)^2(-d\tau^2 + dx^2 + dy^2 + dz^2). \quad (1)$$

We consider the evolution of a real global field $\Phi_a$, $a = 1, \ldots, 4$, whose dynamics are specified by the Lagrangian density

$$\mathcal{L} = -\frac{1}{2}\partial_\mu\Phi \cdot \partial^\mu\Phi - V_0(|\Phi|^2 - \phi^2(T))^2. \quad (2)$$

The function $\phi^2(T)$ is such that the field $\Phi$ undergoes a phase transition at $T = T_c$ from a symmetric phase $(T > T_c)$ to a broken symmetry phase $(T < T_c)$. The symmetric phase corresponds to $\phi^2(T) < 0$, giving the field an $O(4)$ symmetry. When $\phi^2(T) > 0$, the field falls into the broken symmetry phase with $O(3)$ symmetry. The vacuum manifold becomes $S^3 = O(4)/O(3)$. For the range of dynamics we consider, we assume $T \ll T_c$ and $\phi^2(T) = \phi_0^2$, a constant.

Textures form as horizons meet, which has the length scale of Hubble's constant $H$. The natural length scale for textures is the inverse of the radial excitation mass of the field. It is at this scale that we expect a collapsing "knotted" texture to unwind [12]. Borrill [13] has pointed out that for a GUT scale phase transition, these scales differ by $10^{50}$ at the time of decoupling. Any attempt to numerically simulate this model can not hope to cover this range of scales and an approximation must be made.

In this work, we follow [8] and use the non-linear $\sigma$ model (NLSM). This takes the field's mass to infinity, forcing $\Phi$ to *always* be on the vacuum manifold.

The NLSM model is obtained when the Lagrangian density is rewritten as

$$\mathcal{L} = -\frac{1}{2}\partial_\mu \Phi \cdot \partial^\mu \Phi - \lambda(|\Phi|^2 - 1)^2, \tag{3}$$

where the constraint that the field stay on the vacuum manifold is now imposed via the Lagrange multiplier $\lambda$ (the field rescaling $\Phi_a \to \Phi_a/\phi_0$ is assumed). Variation of this Lagrangian leads to the equations of motion

$$\ddot{\Phi}_a + 2\frac{\dot{a}}{a}\dot{\Phi}_a - \nabla^2 \Phi_a = \left[|\nabla \Phi|^2 - |\dot{\Phi}|^2\right] \Phi_a. \tag{4}$$

To investigate this theory via computer simulation, we simulate the order field evolution on a discrete cubic lattice. We map $\Phi^a(\mathbf{x}, t) \to \Phi^a_{i,n}$, where $i$ labels the spatial cells and $n$ is the temporal index. Without loss of generality, we take the grid spacing $\delta x = \delta y = \delta y = 1$ and set the scale via our choice of the time interval $\delta \tau$. Varying a discrete version of Eq. (4) [8], we get the equations of motion

$$\Phi^a_{i,n+1} = \lambda_i \Phi^a_{i,n} + \delta \Phi^a_{i,n} \tag{5}$$

where

$$\delta \Phi^a_{i,n} = \Phi^a - (a^2_{n-\frac{1}{2}}/a^2_{n+\frac{1}{2}})\Phi^a_{n-1} + (a^2_n/a^2_{n+\frac{1}{2}})\nabla^2 \Phi^a_{i,n} \delta \tau^2. \tag{6}$$

($a^2_{n\pm\frac{1}{2}}$ denotes $a^2(\tau_n \pm \frac{1}{2}\delta\tau)$.) $\lambda_i$ is chosen to enforce the constraint $|\Phi_{i,n+1}|^2 = 1$. Squaring Eq. (5) and solving for $\lambda$ gives:

$$\lambda_i = -\Phi_{i,n} \cdot \delta \Phi_{i,n} \pm \sqrt{1 - |\delta \Phi_{i,n}|^2 + |\Phi_{i,n} \cdot \delta \Phi_{i,n}|^2} \tag{7}$$

Expect for the case mentioned below, we always take the positive square root.

The NLSM implementation provides no method by which a knotted configuration unwinds. Since $\Phi$ is constrained to the vacuum manifold, the equations of motion no longer allow for unwinding. In effect, we have assumed that the spatial scale for unwindings is smaller than the inter-cell distance. If we can recognize a wound configuration once it has collapsed to be comparable to the cell size, we may add an unwinding mechanism "by hand". For each spatial cell, we define an alignment

$$\alpha_i = \frac{1}{6}\sum_j \Phi_i \cdot \Phi_j, \tag{8}$$

where the sum on $j$ runs over the $i^{\text{th}}$ cell's six nearest neighbors. A cell is counter-aligned if $\alpha_i < 0$, when it has become completely wound on the scale of the cell spacing. This can be thought of as a situation where the topological charge within a $3 \times 3 \times 3$ cube centered around the counter-aligned cell becomes $> 0.5$. The unwinding then proceeds by choosing the negative sign in Eq. (7): the cell is moved across the vacuum manifold to a configuration that removes the topological charge [8]. Eq. (7) is thus modified to become

$$\lambda_i = -\Phi_{i,n} \cdot \delta\Phi_{i,n} + \text{sign}(\alpha_i)\sqrt{1 - |\delta\Phi_{i,n}|^2 + |\Phi_{i,n} \cdot \delta\Phi_{i,n}|^2} \qquad (9)$$

and the knotted textures are unwound once they shrink below the cell grid scale. The energy density at each point is given by the $\tau\tau$-component of the stress energy tensor:

$$\rho = \frac{1}{2}|\dot{\Phi}|^2 + \frac{1}{2}\partial_i\Phi \cdot \partial^i\Phi \qquad (10)$$

and we use the obvious discretised version in our analysis.

## III. Numerical Results

We use the above algorithm to investigate the distribution of the energy density during the field evolution from the time of decoupling onward. Since the decoupling time is long after the time of the phase transition, we consider field evolution starting from an initially smooth field configuration. This is achieved by initializing each point on the simulation grid to be uniformly distributed on the vacuum manifold and allowing the equations of motion to evolve the field until it "settles down". Initially, the $\alpha_i$'s will be uniformly distributed in $[-1, 1]$. If the unwinding mechanism (Eq. (9) instead of Eq. (7)) were used, it would lead to an unstable field evolution. Instead, we use Eq. (7), choosing *strictly* the positive square root and thus keeping the gradients small. We continue until the total number of counter-aligned cells stops decreasing. Then the unwinding mechanism is enabled and Eq. (9) is used to evolve the field. This procedure allows the field evolution algorithm to generate the required smooth initial conditions. The late time behaviour of the field is insensitive to the precise time that the unwinding is enabled.

We have run a simulation of a $64^3$ element spatial cube for 32 conformal time units ($c = 1$, $\delta x = 1$). Since periodic boundary conditions are used, this is the maximum length of time the simulation can run for and still be viewed as physical. We took $\delta\tau = 0.2$. The evolution was insensitive to taking $\delta\tau \to \delta\tau/2$. The evolution was also insensitive to using a $32^3$ elements spatial cube. The simulation became stable for $\tau > 10$, at which time the unwinding was enabled.

The simulation shows the expected behaviour as regions of above average energy density form, shrink in spatial extent while the energy density rises, and then re-expand while the energy density falls back to the average background level. The time scale for this rise and fall is of order $\Delta\tau \sim 3$. We call such a process an *event*. Typically we find the energy density at the center of an event to be at least $4\sigma$ above the average.

Fig. 1 shows the energy density vs cell alignment for the 20 most energetic events. The alignment decreases and the energy increases as the texture becomes increasingly wound. There were three unwinding events ($\alpha < 0$) in this simulation:

the energy density increases by an order of magnitude as the knotted texture unwinds. The rest of the events never become completely wound. The alignment decreases to a minimum $\alpha > 0$ and then starts to increase again. Shortly after the alignment stops decreasing the energy density stops increasing and the event dissipates.

Fig. 2a shows the distribution of minimum alignments for all events. The distribution's median is $\bar{\alpha} \simeq 0.48$. The most likely event is one where the texture never becomes completely knotted. Although the energy density in these partially-wound events is an order of magnitude below the fully knotted configuration, they are so numerous as to dominate the total energy density of the simulation. Fig. 2b shows the energy density integrated over time for all events in the simulation vs the minimum alignment of each event. This distribution's median is $\bar{\alpha}_\rho \simeq 0.44$. The dominant energy contribution in a texture cosmology comes from partially wound configurations, with only minor contributions from unwinding knots.

## IV. Discussion

The Sachs-Wolfe effect [14] relates CMB anisotropy to perturbations in the gravitational potential, and is usually given in terms of an integral over the the time rate of change of the gravitational potential. For texture models, the gravitational potential results from the kinetic energy density of the order field. There does not appear to be a great variation in the spatial profile of the individual events; hence, the contribution of each texture event to the CMB anisotropy should be proportional to the integrated energy density in the event.

Numerical simulations of the time evolution of global textures demonstrate that the integrated energy density is dominated by partially-wound configurations and not by unwinding knots. The gradient energy density due to texture events is centered about textures with alignment $\bar{\alpha}_\rho \simeq 0.44$. This result proves Borrill *et. al.* [11] claim, but now in the more realistic scenario of an expanding universe.

Efforts to obtain observable predictions for specific texture models have either relied on computationally expensive numerical simulations or on simplifying assumptions. The first method suffers from logistical problems in that proper assessment of cosmic variance requires hundreds to thousands of realizations for each choice of model parameters. The second method typically suffers from assumptions (spherical symmetry, uncorrelated placement of subsequent textures, fully wound textures) that may not be well founded. We have demonstrated that any simplified texture model must consider the full distribution of partially-wound events along with the rarer high-energy events. We are currently attempting to derive a statistical description (distribution of shapes, spatial and temporal correlations) of the gravitational potential in texture cosmologies. If a sufficiently compact description can be obtained which retains the distinctive features of a full-blown numerical simulation, significant computational savings may be realized in Monte Carlo simulation of CMB anisotropy for texture models. These techniques may then be incorporated into other analytic models [15, 16] to compare specific texture cosmologies with the observed CMB anisotropy on various angular scales.

## Acknowledgements

We would like to thank Shoba Veeraraghavan for her time and helpful discussions. One of us (NGP) would also like to thank Fred Lang of Catholic University of America for making the funds and resources available to facilitate this work.

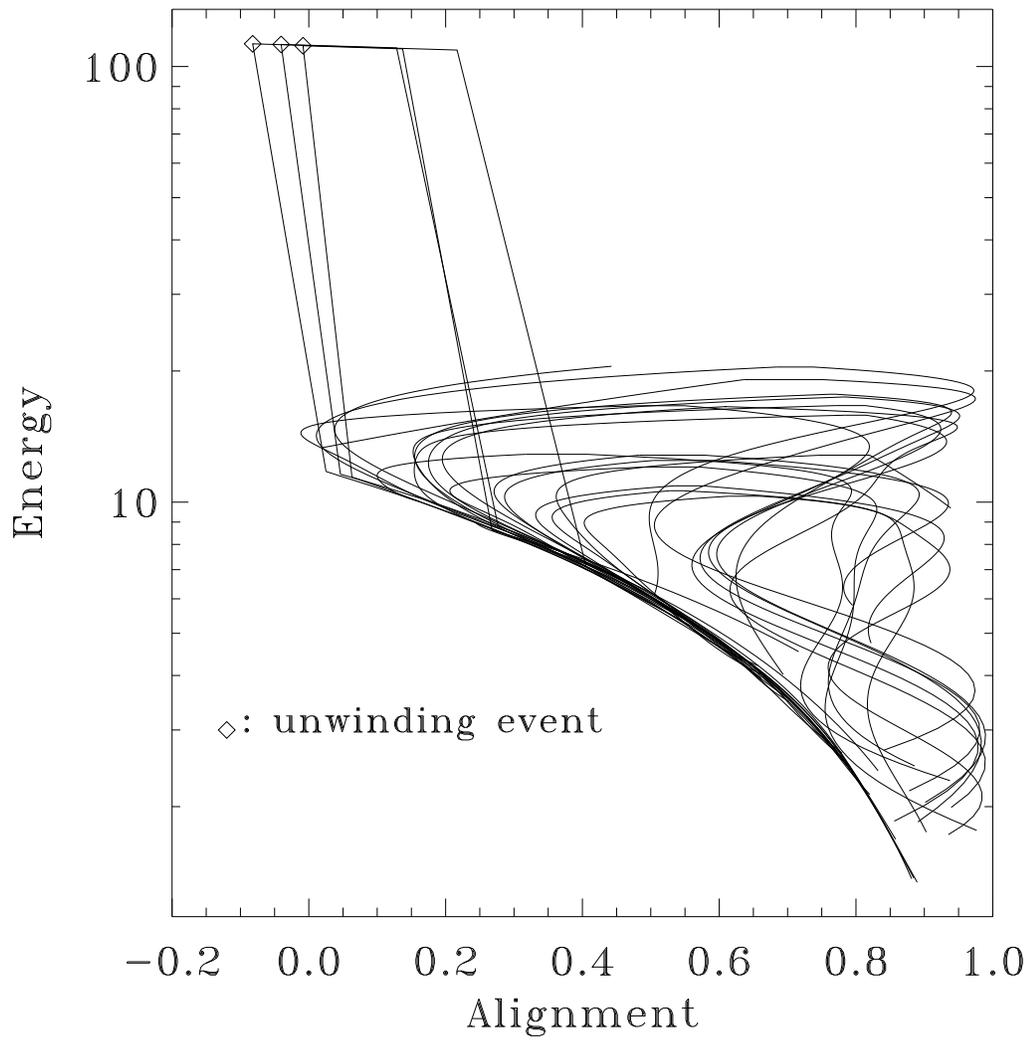

Figure 1: Evolution of texture events viewed in the energy density-alignment plan. Diamonds denote points in the evolution where the alignment was negative (an unwinding event).

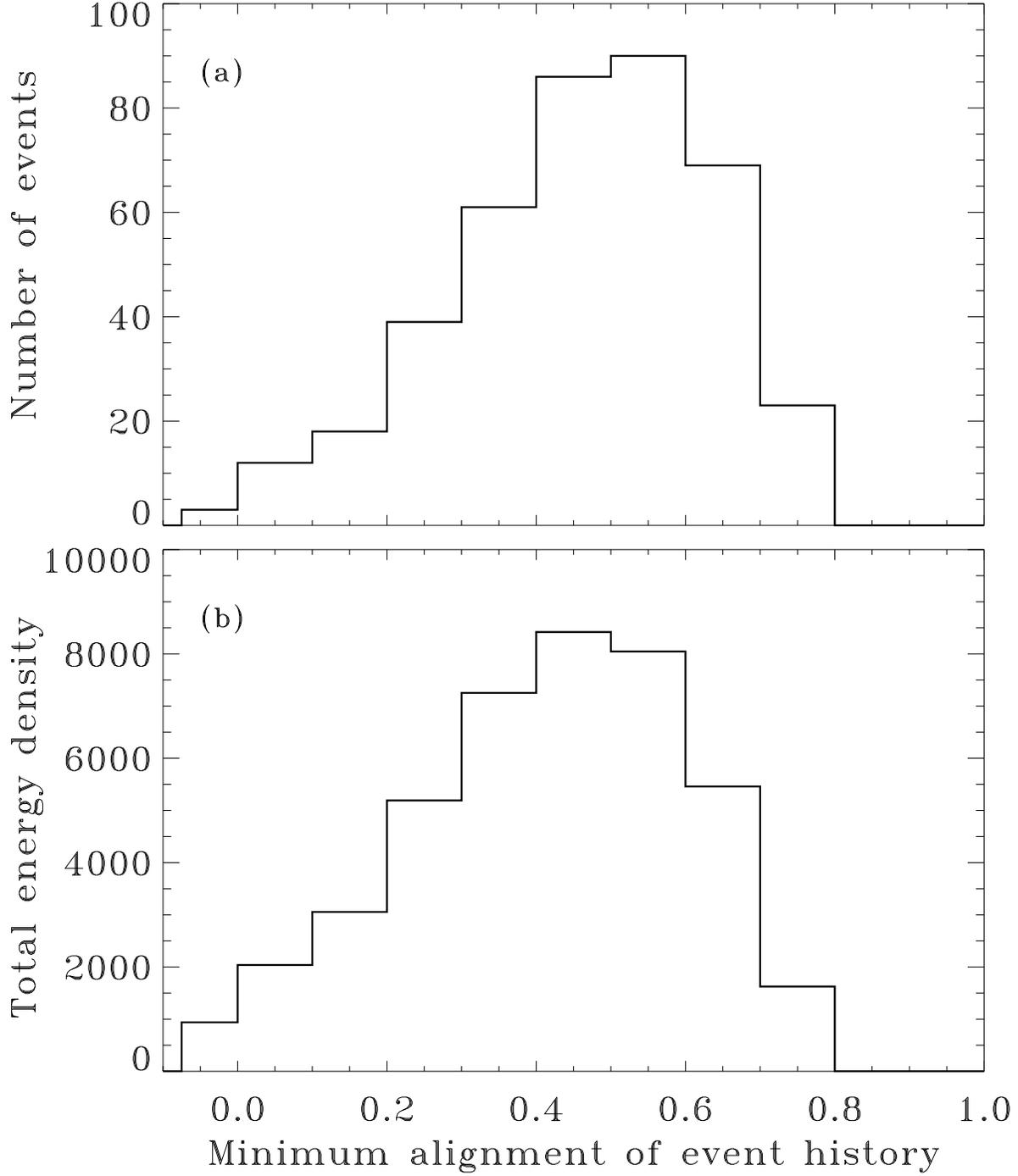

Figure 2: (a) Distribution of minimum alignments of event histories. (b) Energy density (arbitrary units) integrated over time, sorted by the minimum alignment of each event. The median for energy distribution is $\bar{\alpha}_\rho \simeq 0.44$, from texture events that are only partially wound.

10